\documentclass{article}
\usepackage{amsfonts}
\usepackage{spconf,amsmath,graphicx,epsfig}
\usepackage{graphicx}
\usepackage{color}
\usepackage{mathtools}
\usepackage{microtype}
\usepackage{xcolor}
\usepackage{caption}
\usepackage{subcaption}
\usepackage{lineno}
\usepackage{tabularx}
\usepackage{units}

\usepackage{tabularx}

\newcommand{\amy}[1]{\textcolor{black}{#1}}
\newcommand{\alex}[1]{\textcolor{black}{#1}}
\newcommand{\alexx}[1]{\textcolor{black}{#1}}
\newcommand{\huck}[1]{\textcolor{black}{#1}}
\title{Low-Resource Music Genre Classification with \\ Cross-Modal Neural Model Reprogramming}

\name{Yun-Ning Hung$^{1}$, Chao-Han Huck Yang$^{1}$, Pin-Yu Chen$^{2}$, Alexander Lerch$^{1}$}
\address{$^1$ Georgia Institute of Technology, Atlanta, GA, USA\\$^2$IBM Research, Yorktown Heights, NY, USA }

\def\authorname{Y.-N. Hung, C.-H. Yang, P.-Y. Chen, A. Lerch}

\usepackage{hyperref}

\begin{document}

\maketitle
\begin{abstract}
Transfer learning (TL) approaches have shown promising results when handling tasks with limited training data. \amy{However,} \huck{\alexx{considerable} memory and computational resources are often required} \alex{for fine-tuning pre-trained neural networks with target domain data}. In this work, we \huck{introduce a \alex{novel} \alexx{method for} \alex{leveraging}} pre-trained speech models for low-resource music classification based on the concept of \textit{Neural Model Reprogramming (NMR)}. NMR aims at re-purposing a pre-trained model from \amy{a} source domain to a target domain by modifying the \amy{input} of a frozen pre-trained models for cross-modal adaptation. In addition to the known, \amy{input-independent}, reprogramming method, we propose an new reprogramming paradigm: \textit{
Input-dependent NMR}, to increase adaptability to complex input data such as musical audio. 
Experimental results suggest that a neural model pre-trained on large-scale datasets 
can successfully perform music genre classification by using this reprogramming method.
The two  proposed Input-dependent NMR TL methods outperform fine-tuning-based TL methods on a small genre classification dataset. 

\end{abstract}
\section{Introduction}\label{sec:introduction}
\huck{Large-scale datasets are often recognized as one key component in successfully building  powerful deep neural network (DNN) prediction models
~\cite{lecun2015deep}. For example, the ImageNet dataset~\cite{deng2009imagenet} with 14 million samples can be used to train several benchmark image classification systems~\cite{russakovsky2015imagenet} for visual perception. For sequence modeling and language processing tasks, the representation power of models pre-trained on millions of language data such as BERT~\cite{devlin2018bert} 
demonstrates promising results on few-shot tasks and low-resource prediction. In speech and acoustic understanding applications, the performance of acoustic models also benefits considerably from large-scale datasets such as human speech commands~\cite{warden2018speech} and audio event prediction datasets (e.g., AudioSet~\cite{gemmeke2017audio}).}

In the music domain, however, the lack of large-scale training data has been a critical problem. This not only hinders the development of data-driven 
approaches
, but also makes \alexx{the investigation of novel deep learning architectures (e.g., transformer-based models) developed in other domains challenging since they often require massive amounts} of training data.
To tackle 
this problem, Transfer Learning (TL) is a popular solution. 
For example, the MusiCNN~\cite{pons2019musicnn} and JukeBox~\cite{dhariwal2020jukebox} models, pre-trained large-scale music datasets, 
have achieved promising performance on several low-resource MIR downstream tasks~\cite{korzeniowski2020mood}. VGGish and the OpenL3, pre-trained on the large-scale audio dataset, AudioSet~\cite{hershey2017cnn, cramer2019look}, also have shown their effectiveness on various music information retrieval (MIR) downstream tasks~\cite{lianguser,koh2021comparison}. 

Although these TL methods \huck{could demonstrate decent performance in downstream tasks,} they suffer from some drawbacks. For example, the learned representation \huck{does not guarantee to contain task-specific information.} Recent results show, e.g., that models pre-trained on music auto-tagging might lack of information for key detection \cite{castellon2021codified}. Moreover, it seems that learned representations might still lead to inferior performance compared to task-specific models designed for the purpose \cite{castellon2021codified}. 
Theoretically, fine-tuning the pre-trained model should allow to solve this problem. However, training large-scale models such as JukeBox or VGGish requires considerable computational resources. 

In this work, we attempt to tackle this problem from a different angle. We take advantage of a newly proposed method called Neural Model Reprogramming (NMR) \cite{elsayed2018adversarial, yang2021voice2series, vinod2020reprogramming}, to adopt a pre-trained model for downstream tasks. NMR is an alternative TL technique that has been confirmed to provide good or even state-of-the-art results~\cite{yang2021voice2series, vinod2020reprogramming} in numerous machine learning tasks. It aims to \huck{re-purpose a (frozen) pre-trained model on a task-specific dataset} with only a small amount of training parameters without modifying the whole model. Unlike the original NMR training scheme, which proposes adding an universal trainable noise directly to the input sequence, \amy{we further propose two novel input-dependent methods to learn sample-dependent information from different input signals and improve the NMR training strategy on musical data. }

We investigate the proposed method on two \alex{models pre-trained on} audio and speech, respectively: Audio Spectrogram Transformer (AST)~\cite{gong2021ast} and SpeechATT~\cite{de2018neural}, for the task of music genre classification. By using the proposed methods, we achieve new state-of-the-art results on the small music genre classification dataset, GTZAN \cite{sturm2013gtzan}.


%

\vspace{-0.8\baselineskip}

\section{Related Work}\label{sec:related_work}

\begin{figure}
    \begin{subfigure}[b]{\columnwidth}
         \centering
         \includegraphics[width=\columnwidth]{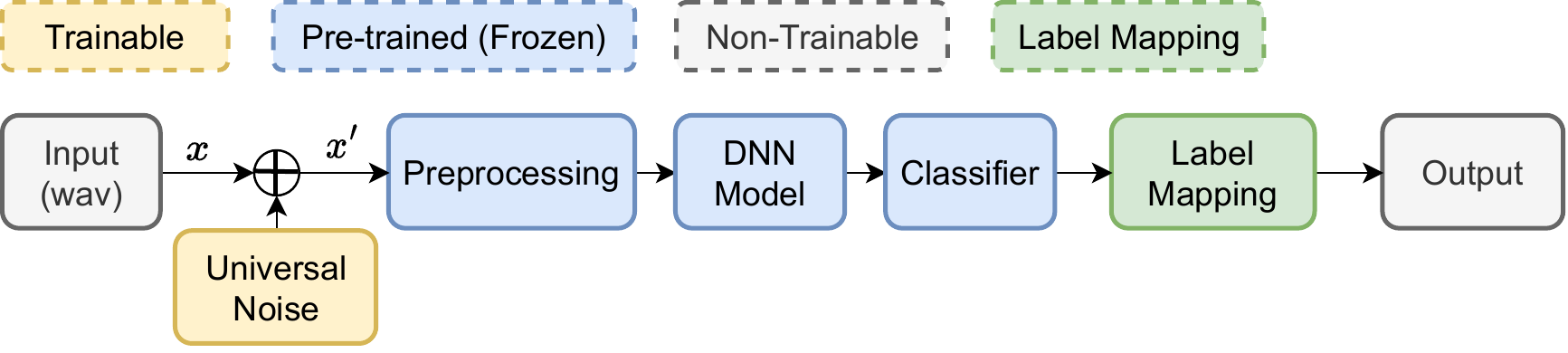}
         \caption{Training pipeline for the baseline input-independent NMR by adding an universal noise to the input waveform (II-NMR).}
         \label{fig:reprogram_noise}
     \end{subfigure}
     \par\bigskip
     \begin{subfigure}[b]{\columnwidth}
         \centering
         \includegraphics[width=\columnwidth]{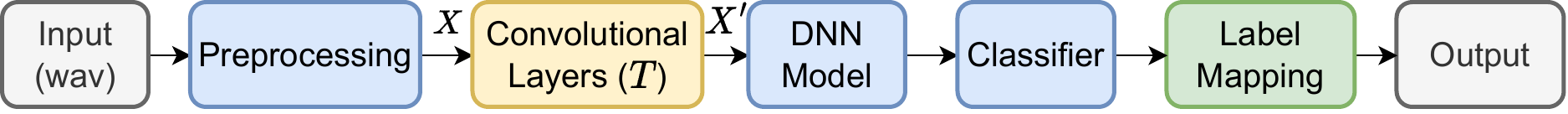}
         \caption{Training pipeline for the proposed input-dependent NMR by transforming the input feature (ID-NMR).}
         \label{fig:reprogram_depend}
     \end{subfigure}
     \par\bigskip
     \begin{subfigure}[b]{\columnwidth}
         \centering
         \includegraphics[width=\columnwidth]{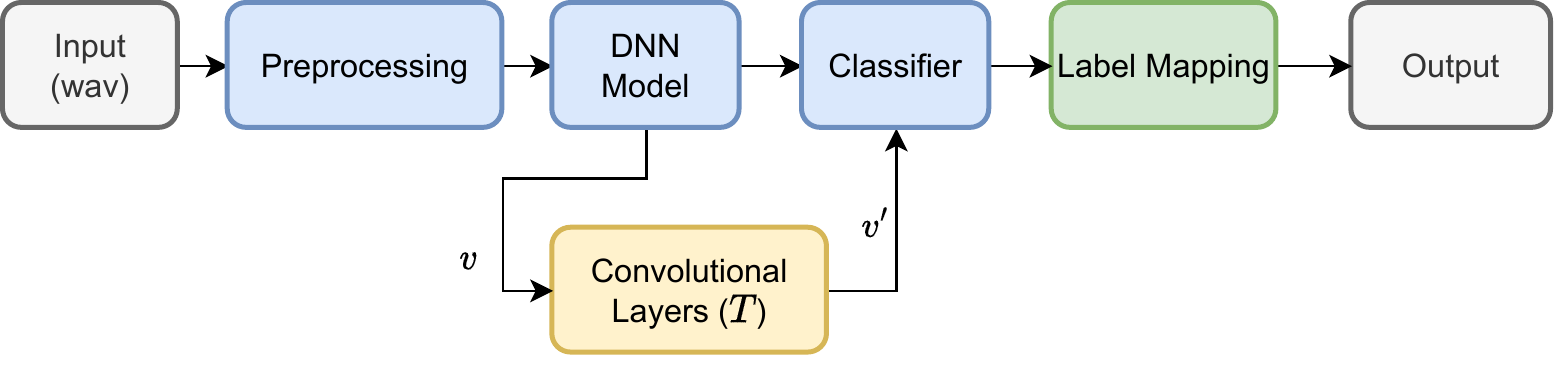}
         \caption{Training pipeline for the proposed input-dependent NMR method with skip connection (IDS-NMR).}
         \label{fig:reprogram_skip}
     \end{subfigure}
    \caption{The overview of the training pipeline for the three NMR methods used in this work.}
    \label{fig:pipeline}
\end{figure}

The concept of NMR is originally related to adversarial attacks \cite{43405}. By applying ``trainable perturbations'' against the loss function of a target model on the input data, an adversarial noise can maliciously manipulate output predictions of neural network based classification model. 
Meanwhile, Elsayed et al.~\cite{elsayed2018adversarial} proposed using this trainable noise to \textit{reprogram} a model pre-trained on a specific ``source'' task (e.g., ImageNet) to a new ``target'' task (e.g., MNIST or CIFAR-10). \huck{After the reprogramming model is successfully trained on the target data, these reprogramming noises can be directly applied to a frozen pre-trained model input during inference time. 
These trainable perturbations can thus be considered a new \textit{program} to empower cross-domain transfer learning toward new tasks. NMR also demonstrates its flexibility in an application-oriented setting~\cite{tsai2020transfer}, such as reprogramming a frozen traffic sign image prediction API into a medical image classifier without knowing the parameters of a source model.}

In addition to image data, the NMR method has been proven effective in various domains. In natural language processing, NMR has been proposed for machine translation and sentiment classification \cite{hambardzumyan2021warp}. 
Vinod et al.\ further explore the possibility of cross-domain reprogramming by adopting NLP models to molecule learning tasks in biochemistry \cite{vinod2020reprogramming}. \huck{In terms of time series processing, the recent advances~\cite{yang2021voice2series,yen2021study} of Voice2Series show an English speech command model could be re-programmed 
to become either a sensor data predictor or a multilingual recognizer on some low-resource language such as Lithuanian with state-of-the-art performance.} 

The aforementioned techniques provide both preliminary empirical findings and theoretical foundations to further motivate us to study NMR techniques toward music applications, addressing  the limitations of insufficient training data and infeasible large-scale neural architectures.

\vspace{-0.4\baselineskip}

\section{Neural Model Reprogramming}
\alexx{This work introduces two new reprogramming methods. While traditional reprogramming simply adds noise to the input of a system as described in Sect.~\ref{ssec:nmr}, our methods transform the input signal non-linearly and utilize intermediate representations to account for the variability and complexity of musical input signals. As shown in Fig.~\ref{fig:pipeline}, all our pre-trained systems include input waveform, preprocessing step to extract a high-level representation, deep neural network, and a classifier to predict the probability of each label. The output of the classifier is then mapped by a many-to-one label mapping layer to map the probabilities of $n$ source classes to a target class. That is, for a target label $y_\mathrm{T} \in Y_\mathrm{T}$, the prediction will be the averaged class predictions ($\{y_1, y_2, ..., y_n\} \in Y_\mathrm{S}$) over the set of source labels from the original pre-trained model assigned to it. All of the steps mentioned above will not be updated during training.}
Instead, a trainable transformation function $\mathcal{H}$ will be added to introduce "noise" to the training pipeline. We will introduce how to add $\mathcal{H}$.

\vspace{-0.5\baselineskip}

\subsection{Input-independent NMR: Baseline Method}\label{ssec:nmr}
\huck{\alex{Since} the applicability of NMR to music signals remains \alex{unexplored}, we first investigate the efficiency and potential of NMR for music signals by establishing a baseline.} We first follow the same training scheme proposed by Yang et al.\ \cite{yang2021voice2series} as one simple baseline of waveform-level reprogramming. As shown in Fig.~\ref{fig:reprogram_noise}, a  trainable universal parameter, $\theta$, is added as a "noise" to transform the whole waveform: $x^{\prime} = \mathcal{H}(x;\theta)\coloneqq x+\theta$, where $x$ and $x^{\prime}$ represent the waveform before and after transformation, respectively.
Since the universal parameter is independent of the input, we refer to this method as \textit{II-NMR}.

\vspace{-0.5\baselineskip}

\subsection{Input-dependent NMR}\label{ssec::ar-input}
Although most previous NMR studies focus on adding a universal transformation on input waveform, we suspect that a more elaborate transformation is required for the music signal. Hence, we propose the following two adjustments.

First, music signals are fundamentally more complex than other audio data (e.g., speech command). To capture complicated harmonic relations, MIR tasks usually rely on perceptually meaningful features, such as Mel-spectrogram, instead of using raw audio. Moreover, the categorization of music is usually not determined by only one or two factors. In genre classification, for example, the same instrumentation or chord progression can appear in completely different music genres. Therefore, we propose that the trainable parameters should be related to features instead of the raw waveform. Moreover, the trainable parameters should be non-linear and ``input-dependent'' to capture the complex musical features of each training sample.
\footnote{We also experimented on transforming the waveform or adding noise to the features, but these methods resulted in low accuracy.} To achieve these criteria, we add a transformation function ($T$) composed of convolutional layers to transform the original feature, as shown in Fig~\ref{fig:reprogram_depend}. The transformation function will learn how to add "noise" to the input features depending on each input sample. That is, $X^{\prime} = \mathcal{H}(X;\mathrm{w})\coloneqq T(X)$, where $w$ is the parameters of $T$. We refer to this adjustment input-dependent NMR \textit{(ID-NMR)}.

Second, there are few potential drawbacks of directly modifying either waveform or input features. For example, to update parameters of the transformation function, the reprogramming layer still needs the gradient from the whole model during training, which leads to slower training time. Moreover, compared to the raw waveform or input features, middle layers of the pre-trained models sometime represent more high-level information which is more critical for classification. Hence, we propose using skip connections 
to transform the output of the middle layer $v$ of the model, and add the transformed feature $v^{\prime}$ (or noise) to the classifier for classification, as shown in Figure~\ref{fig:reprogram_skip}. By using skip connections, we expect the adaptive information from the input-dependent features from the skip-connection to directly influence the classifier. We will refer to this method later as \textit{IDS-NMR}.

\vspace{-0.5\baselineskip}

\subsection{Source Models}

In this section, we introduce two DNN models that we use in the experiment. To explore the probability of cross domain reprogramming, we choose SpeeechATT and AST models pre-trained on speech and audio data, respectively.\footnote{The preprocessing step, DNN model and classifier all follow the implementation of the original models.}

\textbf{SpeechATT}: \label{sec:speechatt}
we choose SpeechATT as the efficient baseline model since Yang et al. has achieved several state-of-the-art results on time-series data by using the NMR training scheme and the SpeechATT model. 
The model is pre-trained on the Google Speech Command dataset \cite{warden2018speech}, which contains 105,829 utterances of 35 words.  Different than other speech recognition tasks, utterances usually contain very short time frames, so the input of this model is only a one-second audio sample. In music genre classification tasks, the input audio is usually longer than one second. To satisfy the data format of pre-trained speech command model, we 
chunked the input sample \amy{into non-overlapping one-second pieces. 
The final probability of each label is attained by averaging the probabilities over all chunks.}



\textbf{Audio Spectrogram Transformer (AST)}:
In recent years, transformer architectures have been successfully applied to a variety of tasks 
\cite{devlin2018bert, parmar2018image}. 
In this paper, we experiment with the recently proposed AST model \cite{gong2021ast} for reprogramming. 
AST is a purely attention-based model for audio classification with BERT-alike patch-wise feature learning. To further leverage a larger scale of training data, AST is pre-trained on ImageNet \cite{russakovsky2015imagenet} and fine-tuned AudioSet \cite{hershey2017cnn}. 
Although other large-scale models, such as VGGish model \cite{hershey2017cnn}, 
have been commonly used in several MIR downstream tasks \cite{koh2021comparison,lianguser}, 
we choose AST since it has outperformed the VGGish model on AudioSet classification. \amy{Moreover, reprogramming VGG-based architectures has shown worse performance due to the disadvantage of the multi-channel feature aggregation used in VGG-based architectures~\cite{yang2021voice2series}. The input of AST model is an audio recording of length \unit[10]{s}. We apply a similar chunking approach as for SpeechATT: audio 
is chunked into non-overlapping \unit[10]{s} segments. 
}

\vspace{-0.8\baselineskip}

\section{Experiment}\label{sec:experiment}
\vspace{-0.5\baselineskip}

\subsection{Dataset}

We choose GTZAN, a small but popular dataset for genre classification, for the experiment. It contains 10 musical genres, with each genre having 100 audio snippets of \unit[30]{s} length, resulting in a total of 1000 snippets. Since the total hours of this dataset is only around 8 hours, it is suitable to represent the scenario of low-resource training data. We adopt
the ``fault-filtered'' split from \cite{kereliuk2015deep} which addresses some of the reported issues with this dataset \cite{sturm2013gtzan}, resulting in 443 pieces for training, 197 pieces for validation, and 290 pieces for testing. The same dataset split is also used in the MIREX baseline systems~\cite{lianguser}.


\vspace{-0.5\baselineskip}

\subsection{Experimental Setup}

Four baseline methods are included in the experiment for comparison. The first baseline is a simple CNN architecture (\textit{BL-CNN}). There are plenty of ways to build the baseline CNN. We choose ResNet since it has shown its efficiency on audio-tagging tasks~\cite{won2020evaluation}. We simply concatenate four ResNet blocks with kernel size 3 and channel size 50.\footnote{The parameters are roughly tuned to produce the best result.} For the second baseline, 
we simply fine-tune the original pre-trained model with the datasets mentioned above as another common TL method. \amy{We call this method \textit{BL-FT-AST} in short. The above mentioned methods together with our approaches can fit the model on task-specific datasets. } 

Another common transfer learning method is to extract the representation from the last layer of the pre-trained model and train a shallow supervised model for classification. Although this method could not tune the pre-trained models on task-specific datasets, it is included for comparison. Following the work from \cite{castellon2021codified}, the mean pooling representations across time are used to train an one-layer MLPs with 512 hidden units for classification. Except for experimenting on the representation extracted from AST model (\textit{BL-R-AST}), we also include the representation extracted from VGGish~\cite{hershey2017cnn} model (\textit{BL-R-VGGish}) for comparison. We use the Pytorch version of the VGGish model\footnote{\href{https://github.com/harritaylor/torchvggish}{https://github.com/harritaylor/torchvggish}} for this experiment. 

Each setting is trained 100 epochs, and the model that performs the best on the validation set is picked for evaluation. Adam optimizer with 0.0001 learning rate is used for training. Label mapping layers have $n=2$ for SpeechATT and $n=5$ for AST after parameter search.
Each method is trained five times with different random seeds. Following previous works \cite{lianguser,lee2018samplecnn,pons2019musicnn,castellon2021codified} we report the mean and standard deviation of accuracy as the evaluation metrics. For the convolutional layers ($T$), we use a similar architecture as the \textit{BL-CNN}, with a kernel size of 3 and 136 channels, to roughly match the total parameters as the \textit{BL-CNN} model. For the convolutional layers ($T$), we use a similar architecture as the \textit{BL-CNN}, with a kernel size of 3 and 136 channels, to roughly match the total parameters as the \textit{BL-CNN} model.



\begin{table}
\centering
\begin{tabular}{l >{\centering}p{0.16\textwidth}>{\centering\arraybackslash}p{0.16\textwidth}}
  \hline
  Model & SpeechATT$_{\text{audio}}$ & AST$_{\text{audio + vision}}$ \\
  \hline \hline
  II-NMR & 0.399 $\pm$ 0.013 & 0.503 $\pm$ 0.018\\
  ID-NMR & 0.609 $\pm$ 0.005 & 0.802 $\pm$ 0.021\\
  IDS-NMR & 0.657 $\pm$ 0.021 & \textbf{0.828} $\pm$ 0.017\\
  \hline
 \end{tabular}
 \caption{Results on GTZAN dataset comparing different NMR methods and pre-trained models.} 
 \label{tab:gtzan_scores}
\end{table}

\begin{table}
\centering
\begin{tabularx}{\linewidth}{l|c|c}
  \hline
  Method & Pre-trained Model & Accuracy \\
  \hline \hline
  FT-CNN~\cite{pons2019musicnn} & MusiCNN & 0.790 \\
  FT-CNN~\cite{lee2018samplecnn} & SampleCNN & 0.821 \\
  Emb.+SVM~\cite{lianguser} & \cite{pons2019musicnn}+\cite{covington2016deep}  & 0.801 \\
  Probing~\cite{castellon2021codified} & JukeBox & 0.797 \\
  \hline
  BL-CNN & CNN$_{\text{audio}}$ & 0.666 $\pm$ 0.034\\
  BL-R-VGGish & VGGish$_{\text{audio}}$ & 0.765 $\pm$ 0.017\\
  BL-FT-AST & AST$_{\text{audio + vision}}$ & 0.772 $\pm$ 0.013\\
  BL-R-AST & AST$_{\text{audio + vision}}$ &  0.831 $\pm$ 0.005\\
  \hline
  IDS-NMR & AST$_{\text{audio + vision}}$ &  0.828 $\pm$ 0.017 \\
  
  \hline
 \end{tabularx}
 \caption{Result on GTZAN dataset comparing existing works and other baseline methods with the \textit{IDS-NMR} method.}
 \label{tab:sota_gtzan}
\end{table}

\vspace{-0.5\baselineskip}

\section{Result \& Discussion} \label{sec:result}



We then compare the results between different NMR methods. 
\alex{The performance of} different methods by reprogramming the SpeechATT and AST models can be seen in Table.~\ref{tab:gtzan_scores}. We can observe that for both models, directly adding trainable noise (\textit{NMR-Noise}) does not work well. However, our proposed input-dependent methods can largely improve the performance. After adding skip-connection for training, both models achieve obvious performance gain. SpeechATT generally performs worse than AST since the original model is trained only one-second audio. For music genre classification, it is hard to capture context information within such a short audio clip. Moreover, SpeechATT has fewer parameters and is trained with a smaller and music-unrelated dataset.

In the end, we compare baseline methods and existing TL methods with our input-dependent skip connection method. Result can be seen in Table~\ref{tab:sota_gtzan}. We can observe that although our proposed method utilizes a model pre-trained on partially music-related datasets, AudioSet and ImageNet, it outperforms existing models pre-trained on music-specific datasets, such as MSD~\cite{pons2019musicnn, lee2018samplecnn}, and model pre-trained on millions of music data~\cite{castellon2021codified}. Our \textit{IDS-NMR} method also outperforms fine-tuning method (BL-FT-AST). Although our methods perform on par with the representation method (BL-R-AST), our \textit{IDS-NMR} method gives models the flexibility to learn both mid-level and high-level information. As extra findings, our \textit{IDS-NMR} method with jointly ImageNet and AudioSet pre-training could achieve one highest test accuracy (\textbf{85.1\%}) and outperforms AudioSet-only pre-training by 2.8\% within all the random seeds. 
The representation from AST model (\textit{BL-R-AST}) performs better than VGGish model (\textit{BL-R-VGGish}). 
This result suggests that the utilization of the AST model could thus lead to improved results for a variety of MIR applications in the near future, replacing VGGish features which have been commonly used in many music downstream tasks. 


  

\begin{table}
\centering
\begin{tabular}{l >{\centering}p{0.16\textwidth}>{\centering\arraybackslash}p{0.16\textwidth}}
\hline
  Methods & \# Trainable Params & Speed (min:sec) \\
  \hline \hline
  BL-FT & 88,132,063 & -\\
  II-NMR & 160,000 & 10:08\\
  ID-NMR & 234,623 & 10:11\\
  IDS-NMR & 235,680 & 03:50\\
  
  \hline
 \end{tabular}
 \caption{Training speed and number of parameters by using different training methods on SpeechATT and AST model.} 
 \label{tab:speed}
\end{table}

We also compare the number of parameters and training speed for the proposed methods. We use a single GeForce RTX 2080 GPU to train the GTZAN dataset and report how many seconds it takes to complete one epoch. In each epoch, the models visit each training sample once. We can see from Table~\ref{tab:speed} that although with or without skip connection has similar number of trainable parameters for our input-dependent methods, with skip connecting is faster in training. Fine-tuning AST model although can achieve good result, it could not fit on a single GPU for training. 



\vspace{-0.5\baselineskip}

\section{Conclusion}
In this work, we proposed a new TL training scheme to leverage the pre-trained models on downstream tasks. We prob into two pre-trained models, SpeechATT and AST, by using both baseline and improved NMR methods. 
Experiment result on low-resource music genre classification shows that the proposed input-dependent method not only outperforms the fine-tuning transfer learning method but also achieve higher accuracy than existing models pre-trained on music-specific datasets. In the future, we plan to explore the proposed methods on other MIR tasks \amy{which also suffer from data limitation problem}. 
Our implementation will be available at \url{https://github.com/biboamy/music-repro}. 





\newcommand{\BIBdecl}{\setlength{\itemsep}{0.25 em}}

\bibliographystyle{IEEEtran}
\bibliography{ISMIRtemplate}

\appendix

\end{document}